\documentclass[epsfig,twocolumn,floats,showpacs]{revtex4}
\usepackage{graphicx}
\usepackage{dcolumn}
\usepackage{bm}
\usepackage{amssymb}
\usepackage{amsmath}
\usepackage{epsfig}
\usepackage{subfigure}

\begin{document}

\title{Overlaying optical lattices for simulation of complex frustrated antiferromagnets }
\author{Zhi-Xin Chen$^{\text{1}}$}
\author{Han Ma$^{\text{2}}$}
\author{Mo-Han Chen$^{\text{1}}$}
\author{Xiang-Fa Zhou$^{\text{1}}$}
\author{Lixin He$^{\text{1}}$}
\author{Guang-Can Guo$^{\text{1}}$}
\author{Xingxiang Zhou$^{\text{1}}$}
\email{xizhou@ustc.edu.cn}
\author{Yan Chen$^{\text{2}}$}
\email{yanchen99@fudan.edu.cn}
\author{Zheng-Wei Zhou$^{\text{1}}$}
\email{zwzhou@ustc.edu.cn}
\address{$^{\text{1}}$Key Laboratory of Quantum Information,
University of Science and \\ Technology of China, Hefei, Anhui
230026, China\\ $^{\text{2}}$ Department of Physics,
State Key Laboratory of Surface Physics, Laboratory of Advanced Materials and Key Laboratory of Micro and Nano Photonic Structures (Ministry of Education), Fudan
University, Shanghai, 200433, China}

\begin{abstract}

We present design techniques of special optical lattices
that allow quantum simulation of spin frustration in
two-dimensional systems. By carefully overlaying optical
lattices with different periods and orientations, we are able
to adjust the ratio between the nearest-neighbor and
next-nearest-neighbor interaction strengths in a square spin
lattice and realize frustration effects. We show that only laser
beams of a single frequency is required, and the parameter space
reachable in our design is broad enough to study the important
phases in the $J_1$-$J_2$ frustrated Heisenberg model and
checkerboard antiferromagnet model. By using the polarization
spectroscopy for detection, distinct quantum phases and
quantum phase transition points can be characterized straightforwardly.
Our design thus offers a suitable
setup for simulation of frustrated spin systems.


\end{abstract}


\maketitle

\section{Introduction}

Geometrically frustrated systems with strong correlation
have attracted much attention due to the highly nontrivial
interplay between frustration and correlation~\cite{Balents,Diep}.
In such systems, not all pairs of
spins can simultaneously assume their lowest energy configuration,
and the ground state usually has a large number of degeneracies.
In particular, frustration in
quantum antiferromagnets may cause certain types of magnetically
disordered quantum phases, including the resonating
valence bond spin liquid state~\cite{Anderson} and the valence
bond crystal state~\cite{VBC}.
As an exotic state of quantum matter, the spin liquid in quantum frustrated antiferromagnets is a
topic of great interest in condensed matter physics. In such states, the local moments do not
form ordering down to the very low temperatures despite of strong antiferromagnetic couplings.
It has been proposed that these exotic states are closely related to the mechanism of
high-T$_c$ superconductivity~\cite{Anderson}. For instance,
the quantum spin liquid state could become unconventional superconducting state when the charge carriers are introduced.

Despite intense studies over the past several decades, the nature
of spin liquid in the strongly frustrated regime remains poorly
understood. A typical system is the frustrated $J_1$-$J_2$
Heisenberg model on a square lattice. Besides the nearest
neighboring (nn) spin interactions leading to N\'eel state, the
$J_1$-$J_2$ Heisenberg model contains extra next nearest
neighboring (nnn) spin interactions. Frustration may lead to the
destruction of long-range order like N\'eel order, and quantum
disordered phases may show up in the strongly frustrated regime
with appreciable value of $J_2/J_1$. Many candidates for the
ground states have been suggested, such as a columnar spin
dimerized state, a plaquette resonating valence bond phase, and a
columnar spin dimerized state with plaquette-type modulation.
Because of the difficulties in theoretical and experimental
studies of the frustrated spin liquid physics, a quantum simulator
for such systems is highly
valuable\cite{QuantumSimulator,Bloch}.

Recent advance in the manipulations of
ultracold atoms in optical lattice has opened new possibilities
for realizing simulation of many significant strongly
correlated quantum models~\cite{Bloch,Lewenstein}.
Much efforts have been devoted to the simulation of quantum magnetism
in ultracold lattice gases. Recent experiments have shown the evidences of superexchange interactions between two neighboring sites~\cite{Trotzky}.
Very recently, the realization of large scale quantum simulator of frustrated magnetism in
triangular optical lattices has been achieved~\cite{Struck}.
As for the simulation of the geometrically frustrated antiferromagnets,
an effective manipulating method is highly demanded to reach the strongly
frustrated regime of the $J_1$-$J_2$ Heisenberg model where the exotic quantum spin liquid states may appear
around $J_1/J_2 \sim 0.5$.
However, it is technically difficult to achieve the interesting physical regime.
As we know, the amplitude of intersite tunneling rate decreases exponentially with their distance
and thus the nnn tunneling rate is much smaller than the nn tunneling rate in a normal square lattice.
According to the mechanism of superexchange interactions, the value of $J_2/J_1$ is much more suppressed.

Another outstanding difficulty in exploring magnetic systems in cold atoms is the detection
of quantum many-body states. Various detection methods have been proposed during the past years.
The noise correlations techniques~\cite{noise1,noise2}
may detect density-density correlations, which corresponds to the spin-spin correlations.
Analogous to the neutron diffraction in condensed matter systems,
Bragg scattering spectroscopy~\cite{Bragg} may probe the static spin structure factor
of the N\'eel state. Single-lattice
detectors~\cite{singlelattice} may measure magnetic ordering, but is still a kind of destructive technique.
Recently, the polarization spectroscopy
technology \cite{Eckert} was proposed to detect magnetic correlations including ferromagnetic
and antiferromagnetic but also more exotic spin ordering by tuning the parameters of polarized light,
which makes it a valuable tool for quantum states measurement.

Engineering of optical lattices has proven very versatile
as shown in previous studies~\cite{Santos,Duan} where suitable
laser beam configurations have been employed to produce
sophisticated optical lattices. In this work, we push the idea
further by overlaying optical lattices with different periods
and orientations to realize spin frustration effects in
two-dimensional square lattices.
Filling such a carefully designed optical lattice with ultracold fermionic atoms,
we can then manipulate spin-spin exchange interaction
between two next-nearest neighboring atoms in optical lattice by adjusting the optical potentials.
We calculate the tunneling rate in the square lattice and find that
the nnn tunneling rate could be comparable to the nn tunneling rate.
It turns out that the $J_1$-$J_2$ Heisenberg model in the strongly frustrated regime could be realized in which the
exotic spin liquid states may exist. Moreover, we can simulate other geometrically frustrated lattice like
the checkerboard antiferromagnet model.
As for the detection of quantum many-body states, we utilize the polarization spectroscopy
technology~\cite{Eckert} to study the spatially resolved spin-spin correlation functions.
For $J_1$-$J_2$ Heisenberg model, we demonstrate that this promising method can be used to determine
the quantum phase transition points straightforwardly and characterize distinct quantum phases. In particular,
one-to-one correspondence between the extreme points of noise signal and
quantum phase transition points has been achieved.

The rest of the paper is organized as follows. In Sec.
II, we implement the two-dimensional frustrated optical lattice using a special optical engineering technique.
In Sec. III, simulations of the $J_1$-$J_2$ Heisenberg model and the checkerboard antiferromagnet model
are introduced.
In Sec. IV, we employ the polarization spectroscopy
technology to detect the quantum many-body states. Finally Sec. V gives a summary.

\section{the two-dimensional frustrated optical lattice}

The Heisenberg spin interactions between neighboring ultracold atoms in an
optical lattice arise due to the intersite virtual tunneling of
atoms~\cite{Duan, Kuklov}. Consequently, the spin-spin exchange interaction
depends on the intersite tunneling rate $t$ as well as the on-site Coulomb interaction $U$ via
the superexchange mechanism, $J\sim t^2/U$.
To implement a $J_1$-$J_2$ Heisenberg model in the strongly
frustrated regime, the crucial procedure is to design a special two-dimensional optical
lattice in which the nnn tunneling rate $t_2$ can be comparable
to or even larger than
the nn tunneling rate $t_1$. It is hard to achieve by creating ordinary
optical lattices using laser beams of the same wave number $k$ as
is required to trap atoms in a certain state.
However, by properly designing several optical
lattices with laser beams traveling in different directions and
superimposing them elaborately, we can tune the effective potential
barriers along different directions and realize such a frustrated optical
lattice using laser beams with the same wave number $k$.

We start by creating a normal two-dimensional square lattice. First, we use a
strong standing-wave field in the $z$ direction to completely
suppress the tunneling of atoms along the $z$ direction. Then,
as shown in Fig. 1(a), we shine a pair of blue-detuned interfering traveling
laser beams of wave number $k$ in the $z$-$x$ plane, each with an angle of $\pi/4$ with respect to the
$x$-$y$ plane. These two laser beams co-propagate in the $z$ direction
and counter-propagate in the $x$ direction. Together with the strong
trapping field in the $z$ direction, it creates a periodic potential
along the $x$ direction in the $x$-$y$ plane. Similarly, as shown in Fig. 1(a),
we shine one more pair of blue-detuned
interfering laser beams in the $y$-$z$ plane, each with an angle of $\pi/4$
with respect to the $x$-$y$ plane. This creates a periodic potential along
the $y$ direction in the $x$-$y$ plane. The net effect of these trapping lasers
is the formation of the following two-dimensional spatially varying potential profile:
\begin{equation}
\begin{split}
V_1(x,y)=& V_1 [sin^2 (\frac {k} {\sqrt{2}}x+\frac {\pi}{2})+sin^2
(\frac {k} {\sqrt{2}}y+\frac {\pi}{2})],
\label{eq:1st_latt}
\end{split}
\end{equation}
where $k$ is the wave vector for the laser beams and $V_1$ is proportional to
the dynamical atomic polarizability times the laser intensity. Its
projection into the $x$-$y$ plane is $k/\sqrt{2}$. This potential profile then
produces a square optical lattice with a period
$a_1=\lambda/\sqrt{2}$ where the wavelength $\lambda = 2\pi/k$.
We load the optical lattice such that there is one atom trapped at
each minimum of the optical potential. In such an optical lattice,
since the tunneling rate is sensitive to the potential barrier
height and lattice site separation, it can be easily seen from Fig.
1(b) that the nnn tunneling rate along the diagonal direction is
much smaller than the nn tunneling along the $x$, $y$ directions.
This is because the potential barrier along the diagonal direction
is higher and the distance between the sites in that direction is
also larger.

In order to make the nnn tunneling rate $t_2$ comparable to the nn
tunneling rate $t_1$ and thus realize an appreciable $J_2/J_1$, we must
adjust the potential
barrier heights along both the diagonal and $x$, $y$ directions. For this
purpose, we add
another square lattice along the diagonal direction in the $x$-$y$ plane with
a lattice period $a_2=\lambda/2$. As shown in Fig. 1(c),
this second lattice can be created by applying two pairs of standing-wave
laser beams in the $x$-$y$ plane along the diagonal directions and of the same
wave number $k$ as those used earlier to engineer the potential profile in
Eq. (\ref{eq:1st_latt}). Its
lattice potential has the form
\begin{equation}
\begin{split}
V_2(x,y)=& V_2 [sin^2 (\frac {k} {\sqrt{2}}(x +y)+sin^2 (\frac
{k} {\sqrt{2}}(x-y))].
\end{split}
\end{equation}
To avoid undesired interferences between laser beams, one
can randomize the relative orientation of the polarization of
laser beams generating optical lattice potentials of $V_1$ and
$V_2$~\cite{Santos}. Thus, the total effective two-dimensional
optical potential is a sum of $V_1(x,y)$ and $V_2(x,y)$. Its
profile is plotted in Fig. 1(d) for $V_1=10E_R$ and $V_2=15E_R$
where $E_R=\hbar^2 k^2/2m$ is the atomic recoil energy. The
corresponding lattice configuration is depicted in Fig. 1(e). It
is clear that due to the second lattice we engineered, the
effective potential barriers along $x$, $y$ directions are much
enhanced and the tunneling between nearest neighbors is
suppressed. In contrast, the effective potential barrier along the
diagonal direction is lowered and the tunneling between nnn is
enhanced. Thus, we can tune the ratio $t_2/t_1$ by adjusting the
value of $V_2/V_1$.

\begin{figure}
\centering
 \subfigure[]
{\label{1a}\includegraphics[width=0.45\columnwidth]{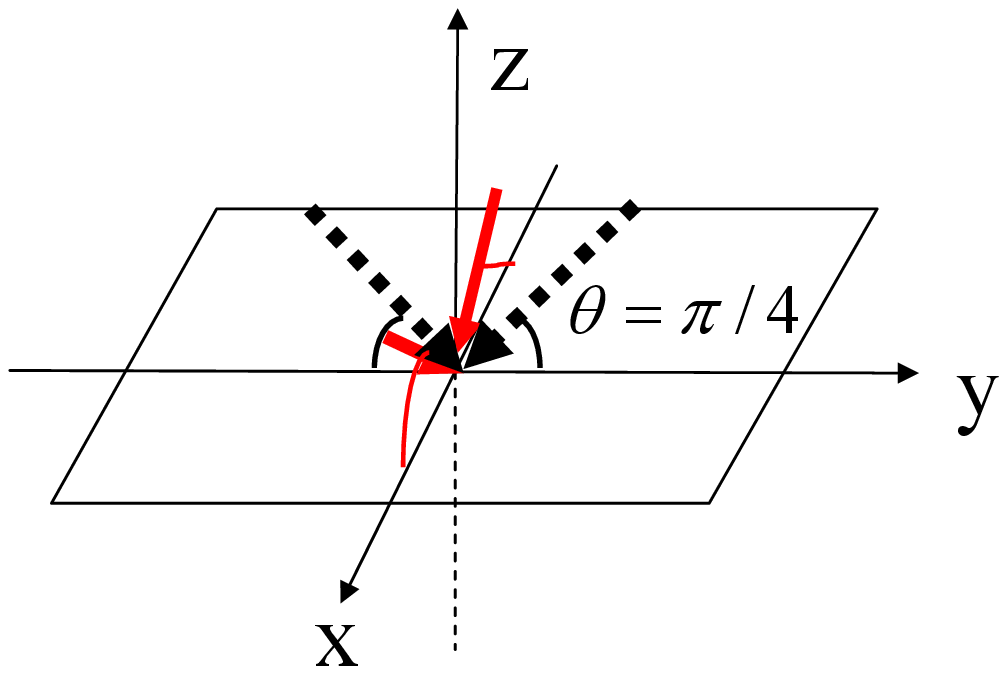}}
 \subfigure[]
{\label{1a}\includegraphics[width=0.45\columnwidth]{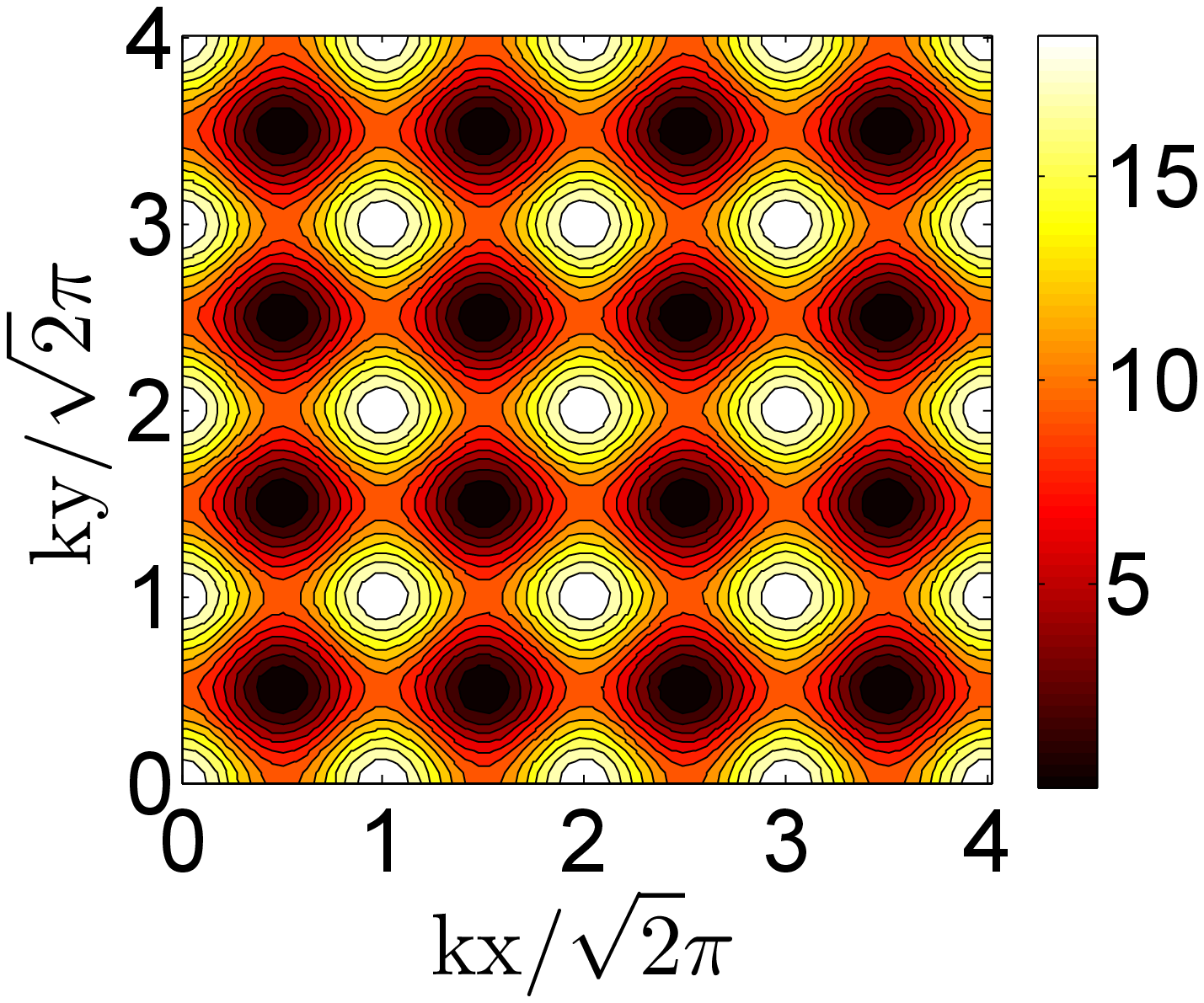}}\\
 \subfigure[]
{\label{1a}\includegraphics[width=0.45\columnwidth]{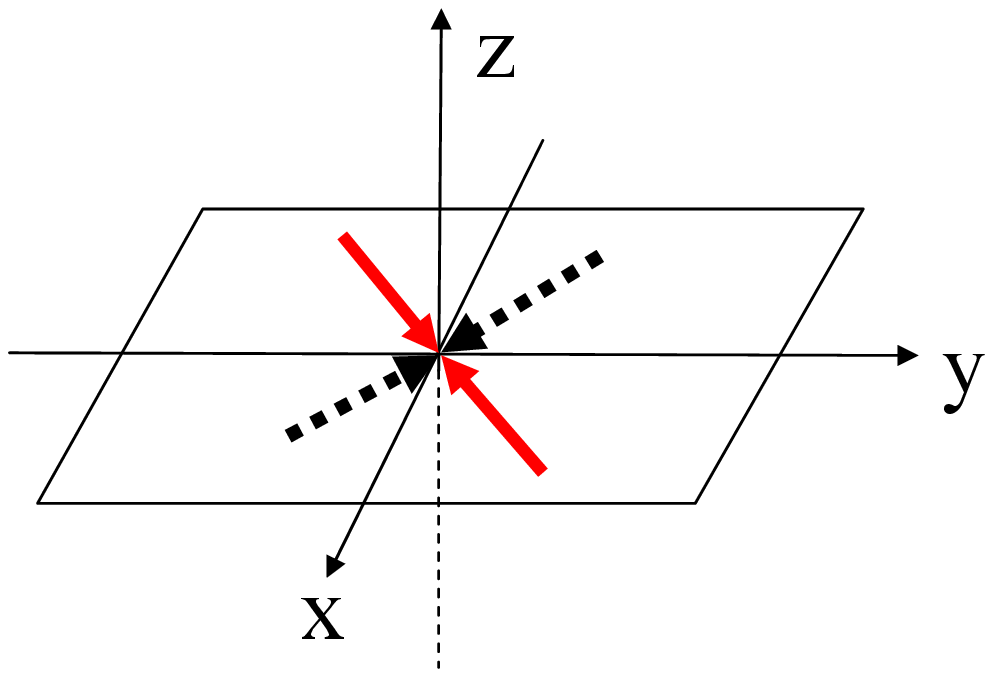}}
\subfigure[]{\label{1b}\includegraphics[width=0.45\columnwidth]{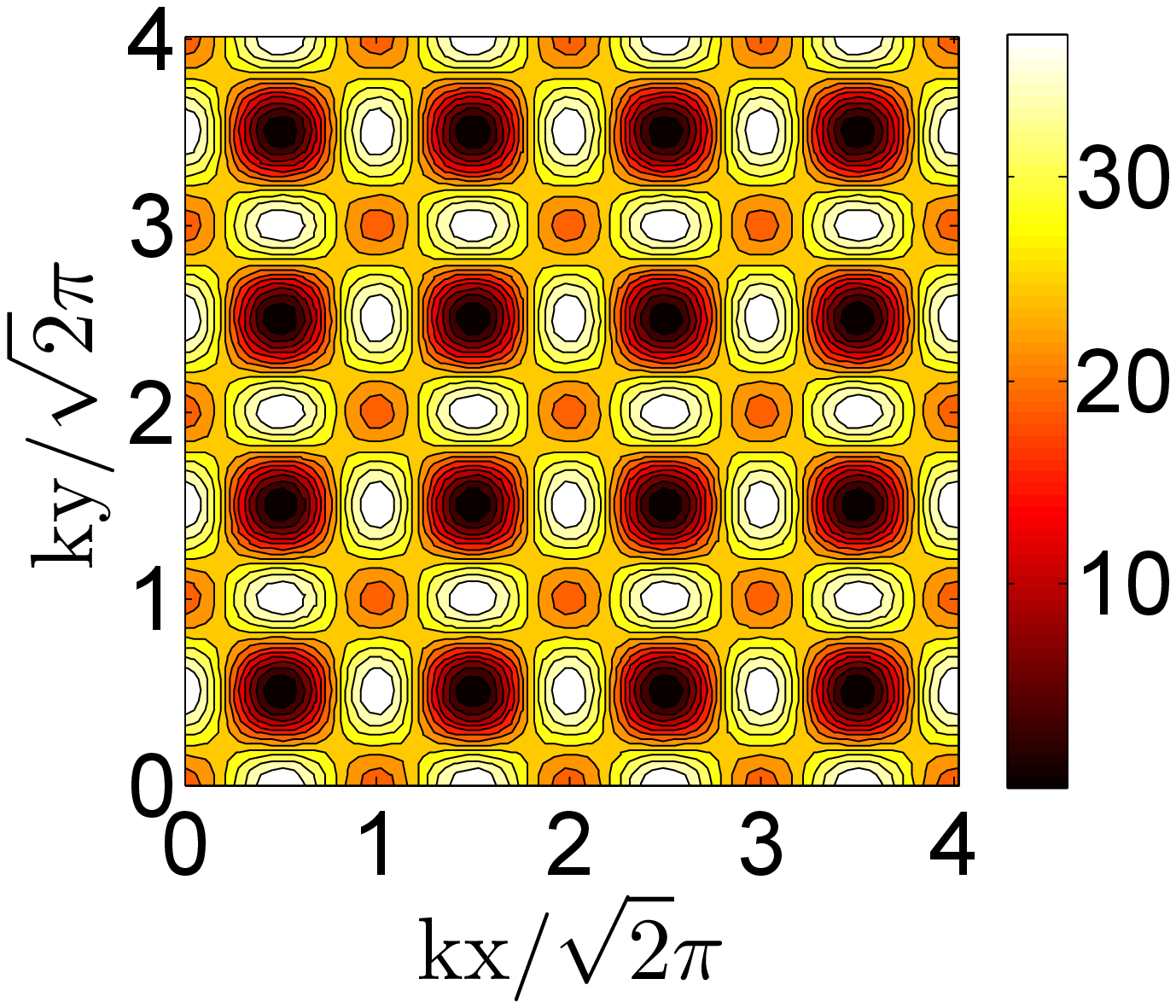}}\\
\subfigure[]{\label{1c}\includegraphics[width=0.40\columnwidth]{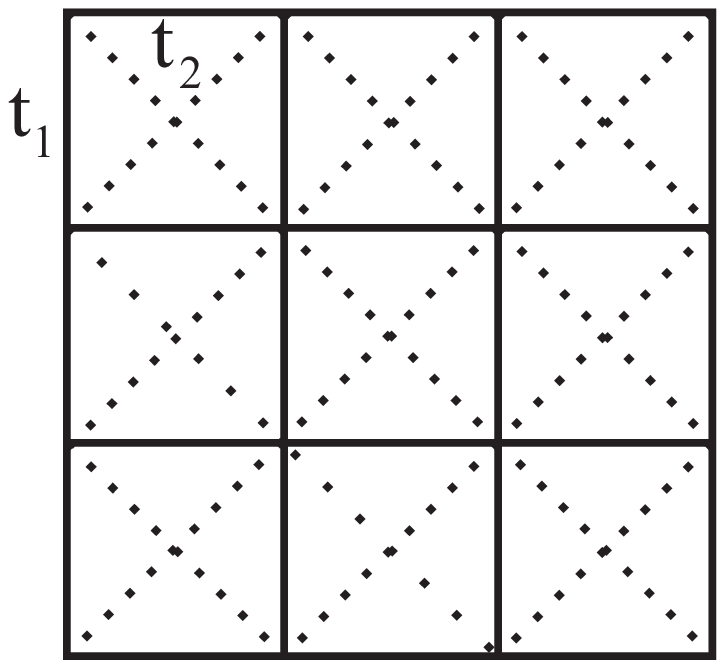}}
\subfigure[]{\label{1d}\includegraphics[width=0.45\columnwidth]{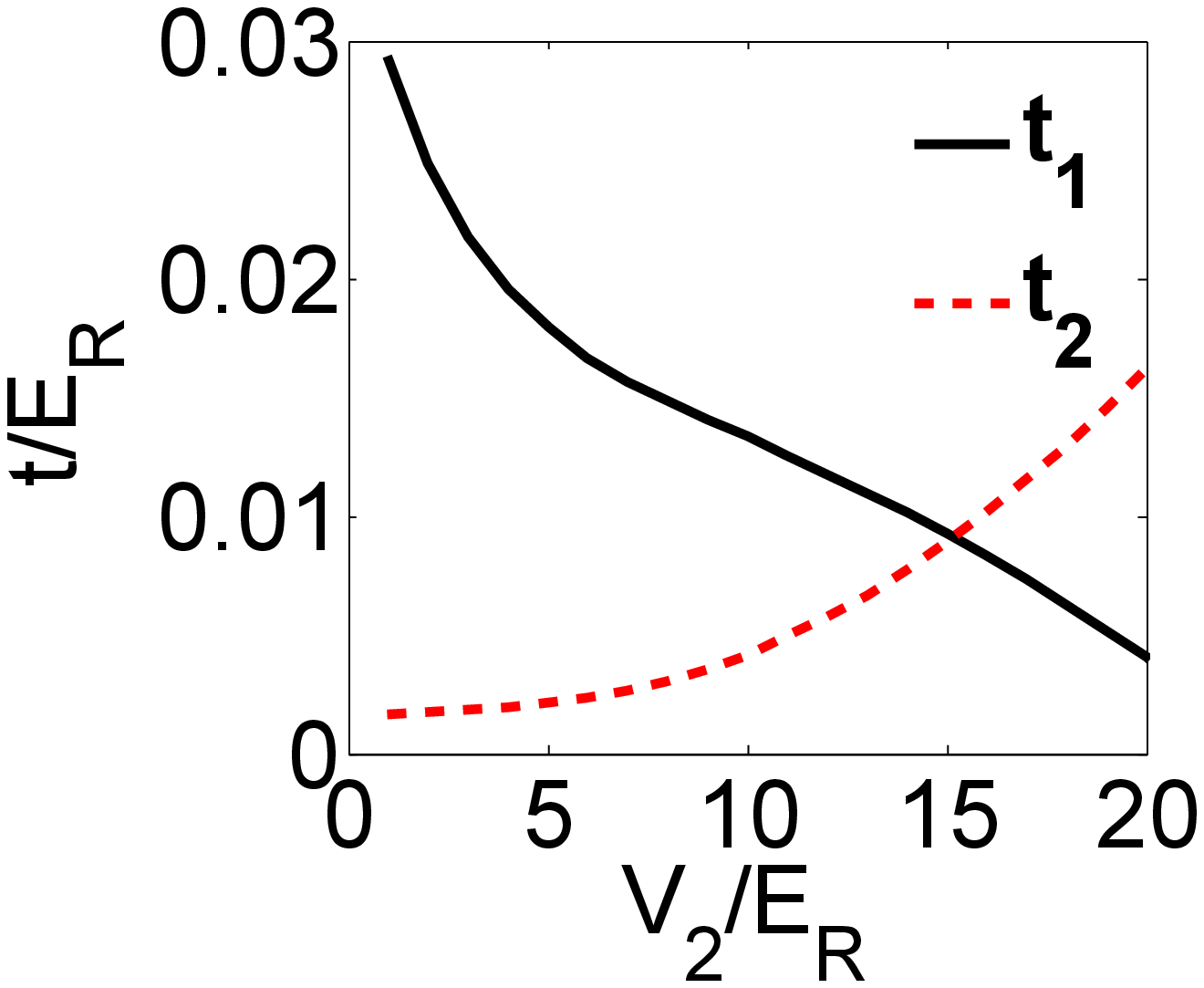}}
\caption{(Color online) (a) Schematic diagram for generating the
interference field $V_1(x,y)$. (b) Normal square lattice pattern induced by the field
$V_1(x,y)$ in unit of $E_R$. (c) Schematic diagram for
generating the interference field $V_2(x,y)$. (d) The square lattice pattern induced by the
field $V_1(x,y)+V_2(x,y)$ in unit of $E_R$. (e) Schematic
diagram of the square lattice with diagonal tunneling. (f) The
nearest-neighboring tunneling $t_1$ (black solid line) and the
next-nearest-neighboring tunneling $t_2$ (red dashed line) as a
function of $V_2$ in unit of $E_R$.}\label{fig1}
\end{figure}

To find out more practical parameter region of $t_2/t_1$ in a
physical system, we numerically calculate these tunneling rates
for an optical lattice loaded with Sodium atoms. The
tunneling rate between site $i$ and site $j$ is shown as,
\begin{equation}
t_{ij}= \langle W_i | H_0 | W_j \rangle\, ,
\end{equation}
where $|W_i \rangle$, $|W_j\rangle$ are the Wannier functions
\cite{marzari97,he01} at sites $i$ and $j$, and $H_0=- \nabla^2/2+
V_1(x,y) + V_2(x,y)$ denotes the single-particle Hamiltonian of the
system. Here we consider the lowest band only. For the Sodium
atoms, $E_R/\hbar=2\pi\times32$ kHz for a blue detuned
$\lambda=514$ nm. We may choose $V_1=10E_R$ and increase $V_2$
from $0$ to $20E_R$ and then plot $t_1$ and $t_2$ in Fig. 1(f). It
is obvious that the value of $t_2$ ($t_1$) increases (decreases)
as $V_2$ increases. Notably, $t_2$ becomes larger than $t_1$ when
$V_2>15E_R$. Therefore, by using a sufficiently large $V_2$, we
can simulate the $J_1$-$J_2$ Heisenberg model in the strongly
frustrated regime. It is worth to mention that we have shown
theoretically the way to simulate an one-dimensional $J_1$-$J_2$
model in concatenated microcavities~\cite{chen}. However, it is
quite difficult to realize a two-dimensional $J_1$-$J_2$ model in
the strongly frustrated regime by employing this method.

Moreover, we can realize a checkerboard lattice by adding a third square
lattice with a lattice period $a_3=\lambda$ along the diagonal
direction in the $x$-$y$ plane. Since the period of this third lattice
is twice of the second lattice and both are oriented in the
same direction,
the diagonal tunnelings in half of the lattice are strongly
suppressed due to the presence of potential maxima, and a checkerboard lattice can
be constructed.
To generate the third lattice, we may adopt a technique similar to that
used for the first simple lattice. In Fig. 2(a), $m$ is one diagonal direction
in the $x$-$y$ plane and $m'$ is the other diagonal direction perpendicular to
$m$. We shine a pair of blue-detuned interfering traveling laser beams of wave number
$k$ in the $m$-$z$ plane, each with an angle $\pi/3$ with respect to the $x$-$y$ plane.
Similarly, another pair of blue-detuned traveling laser beams of the same wave
number are shot in the $m'$-$z$ plane, each with an angle $\pi/3$ with respect to
the $x$-$y$ plane. Obviously, these two pairs of laser beams will form a two-dimensional
lattice along the diagonal directions in the $x$-$y$ plane and its period is determined
by the projected wave number in the
$x$-$y$ plane $k'_\|=k/2$. Specifically, the spatially varying potential profile for the third lattice
is \begin{equation}
\begin{split}
V_3(x,y)=& V_3 [sin^2 (\frac {\sqrt{2}} {4} k(x+y ))+sin^2 (\frac
{\sqrt{2}}{4} k(x-y))].
\end{split}
\end{equation}
For amplitudes $V_1=10E_R$, $V_2=10E_R$ and $V_3=5E_R$, the
superposed potential $V(x,y)= V_1(x,y)+V_2(x,y)+V_3(x,y)$ is
plotted in Fig. 2(b). We can see that there are two sets of plaquettes
in which the potential barrier heights along the diagonal direction
are raised and lowered respectively. Consequently, the diagonal
tunnelings in these two sets of plaquettes are either suppressed or
enhanced, providing us the checkerboard lattice. The configuration of the checkerboard lattice is shown in Fig. 2(c). When
we fix the value of $V_1$ and $V_3$ and adjust the value of $V_2$,
the rate of $t_2/t_1$ will be changed accordingly. When $t_2$ is
comparable to $t_1$, the checkerboard $J_1$-$J_2$ model may reach
the strongly frustrated regime $J_2/J_1 \sim 1$.

\begin{figure}
\centering
\subfigure[]
{\label{2a}\includegraphics[width=0.45\columnwidth]{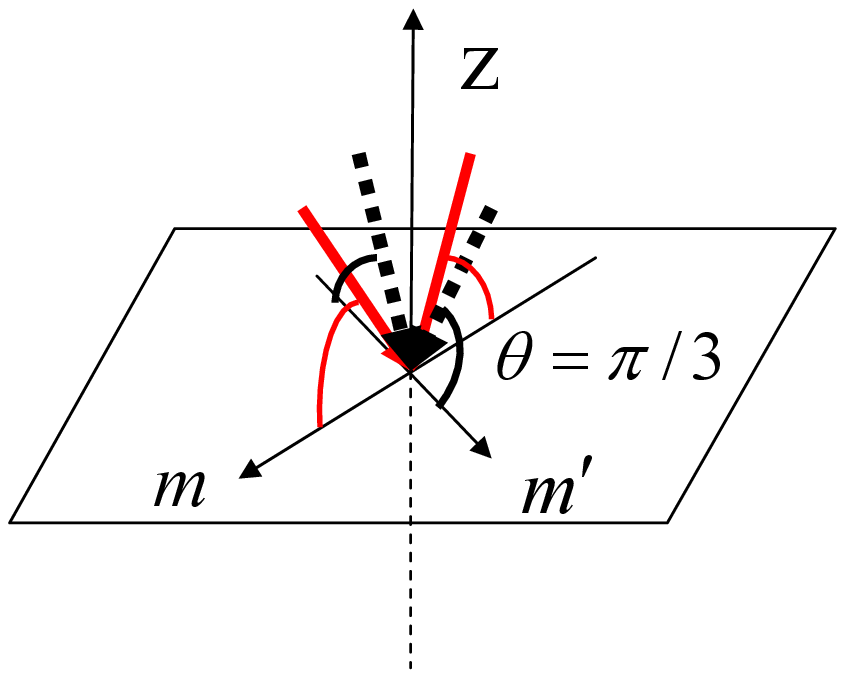}}\\
 \subfigure[]
{\label{2b}\includegraphics[width=0.46\columnwidth]{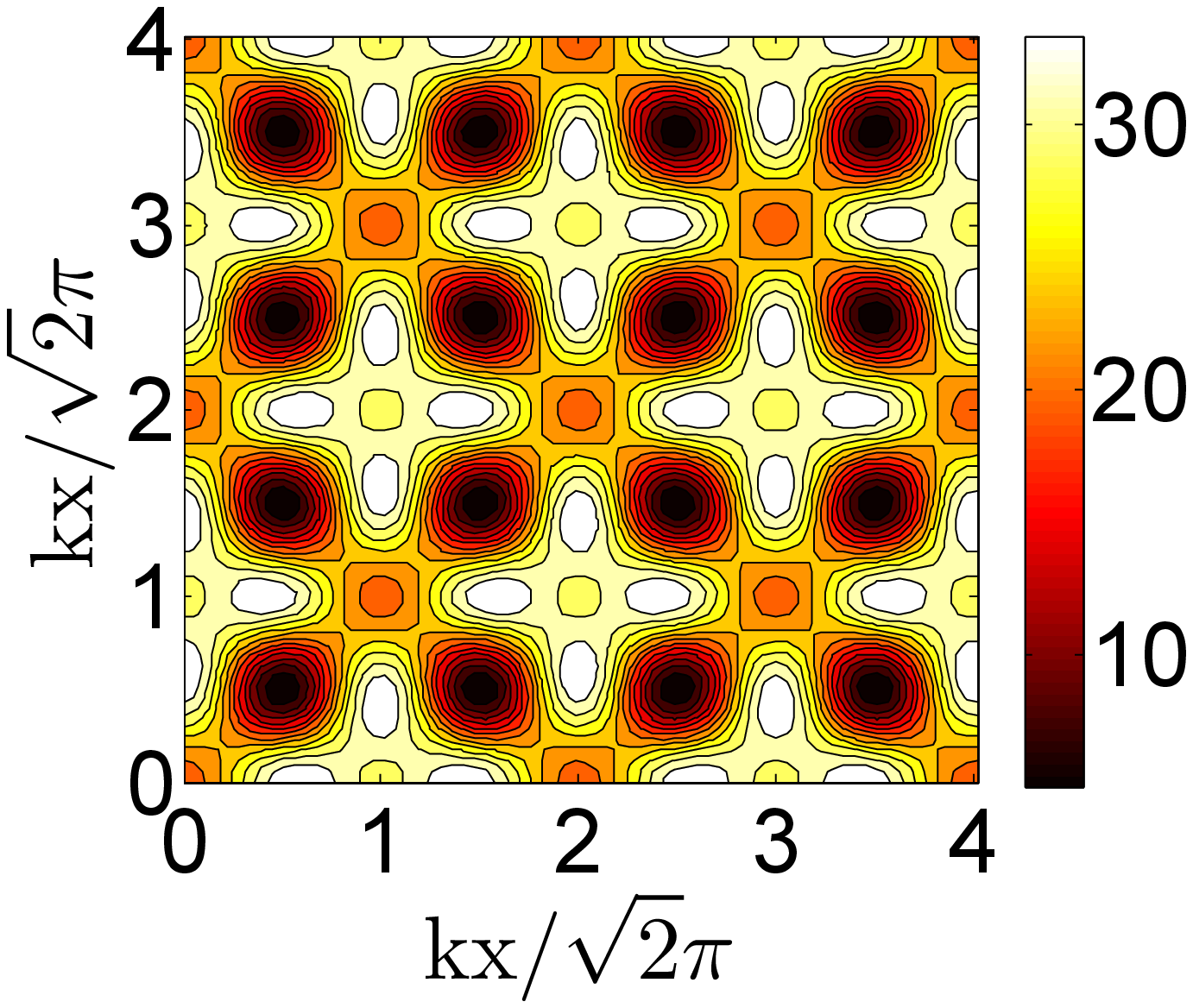}}\\
 \subfigure[]
{\label{2c}\includegraphics[width=0.33\columnwidth]{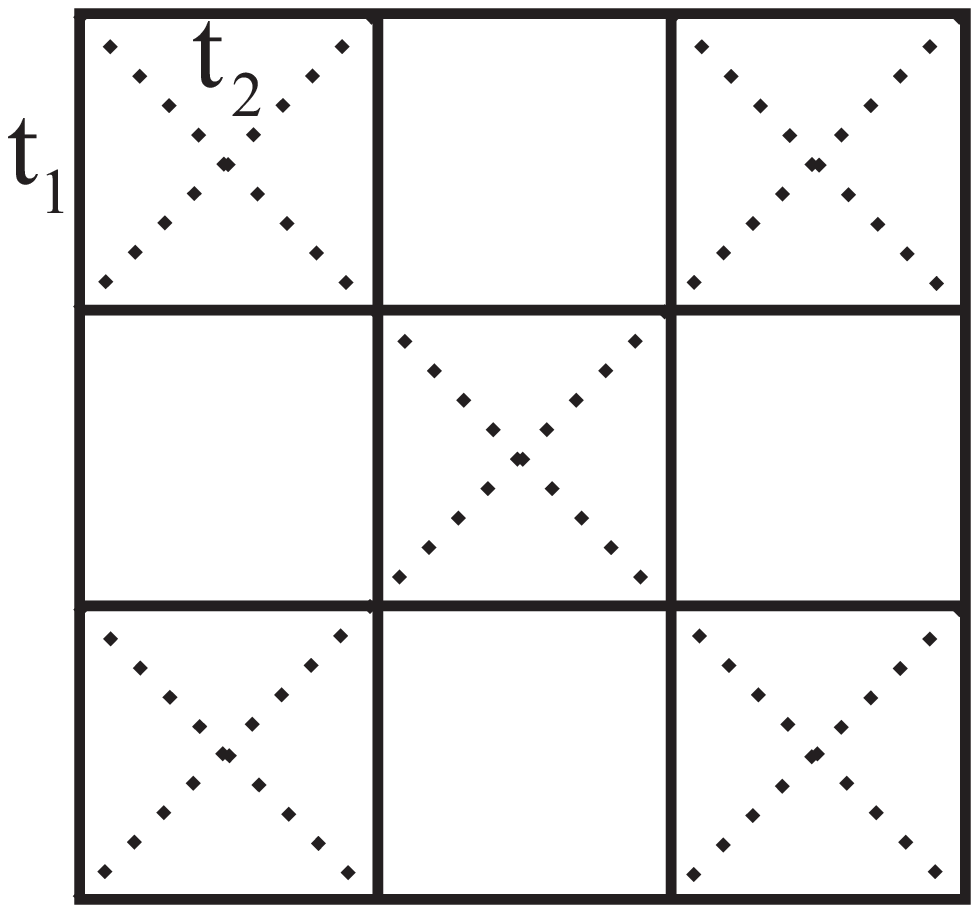}}
\caption{(Color online) (a) Schematic diagram for generating the
interference field $V_3(x,y)$. (b) The optical lattice induced by interference field
$V(x,y)$ in unit of $E_R$. (c) Schematic diagram of a
checkerboard lattice.}\label{fig2}
\end{figure}

\section{the frustrated $J_1$-$J_2$ Heisenberg models}
We can implement the $J_1$-$J_2$ Heisenberg model by using a collection
of ultracold fermionic atoms confined in the special square optical
lattice at integer filling.
In the real experiments, ultracold fermionic atoms have been used to
realize the Mott insulating state~\cite{Jordens, Schneider}. By further lowering the temperatures,
two easily accessible hyperfine states in such system can be used to realize magnetic orders~\cite{Leo,Jordens1}.
Here we consider that the system enters into the Mott insulating state, and then load the ultracold fermionic atoms
onto the optical lattice so that there is only one atom per lattice site. The two hyperfine states
represent the effective spin states
$|\sigma\rangle=\mid\uparrow\rangle$ and $\mid\downarrow\rangle$. We
assume that the optical potential has no state dependence and the
atoms are confined to the lowest Bloch band. The model Hamiltonian can be written as,
\begin{equation}
\begin{split}
H=&-t_1\sum_{\langle i,j\rangle\sigma}(c^\dagger_{i\sigma}
c_{j\sigma}+h.c.)\\
&-t_2\sum_{\langle\langle
i,j\rangle\rangle\sigma}(c^\dagger_{i\sigma}
c_{j\sigma}+h.c.)+U\sum_i n_{i\uparrow} n_{i\downarrow},
\label{eq:Hubbard}
\end{split}
\end{equation}
where $c_{i\sigma}$ is the Fermion annihilation operator for the
atom on site $i$, $n_{i\sigma}=c^\dagger_{i\sigma} c_{i\sigma}$,
$t_1$ and $t_2$ are respectively the tunneling matrix elements between two nn sites
$\langle i,j\rangle$ and two nnn sites
$\langle\langle i,j\rangle\rangle$, and $U$ denotes the on-site repulsive
interaction between the atoms. The interaction $U$ can be adjusted
by Feshbach resonances method. In the Mott insulator regime, we have $t_1$, $t_2$ $\ll$ $U$.
By treating the tunneling as weak perturbation and using the Schriffer-Wolff
transformation, it can be shown that Eq. (\ref{eq:Hubbard}) is
equivalent to the following $J_1$-$J_2$ Heisenberg model Hamiltonian,
\begin{equation}
H=J_1\sum_{\langle i,j\rangle}S_i \cdot S_j+J_2\sum_{\langle\langle
i,j\rangle\rangle}S_i \cdot S_j.
\end{equation}
where $J_i=4 t_i ^2/U$ (i=1,2), $J_1$ and $J_2$ correspond to the
nn and nnn exchange couplings, respectively. (see Fig.1c)
Here we ignore the higher order corrections for the coefficient $J_i$, in which the
dominant item is proportional to $t_i^4/U^3$ \cite{Nevidomskyy}
and can be suppressed greatly as $U/t_i$ grows.

Due to the competition between $J_1$ and $J_2$, the frustrated $J_1$-$J_2$ antiferromagnet has certain limiting cases.
At $J_2/J_1 \ll 1$, the model is reduced to the Heisenberg model on a square lattice.
At $J_2/J_1 \sim 1$, the system has a collinear antiferromagnetic
quasi-long range order. In the strongly
frustrated regime $J_2/J_1 \sim 0.5$, there are several candidates of possible
spin liquid states, but there is still lack of general consensus.
According to previous analytical and numerical studies~\cite{Capriotti, Singh, Sushkov},
there might be three quantum critical points around $J_2/J_1\sim 0.38$, $0.5$, and $0.6$, respectively.
In particular, the simple N\'eel state corresponds to the parameter region $J_2/J_1 < 0.38$, the
simple columnar dimerized spin liquid shows up for $0.38< J_2/J_1 <0.5$, the columnar
dimerized spin liquid with plaquette type modulation may exist in the parameter region $0.5< J_2/J_1 <0.6$, and
the collinear state may appear when $J_2/J_1>0.6$.
As illustrated in Fig. 1d, we observe that $t_2/t_1$ or $J_2/J_1$ increases progressively as we increases $V_2$. It is clear that
the strongly frustrated regime of $J_1$-$J_2$ model can be reached.
This model also exhibits rich dynamical behaviors which makes it an ideal testing ground for the study of quantum phase transitions.
Once the quantum many-body states could be faithfully probed, our proposed quantum simulator may resolve
many outstanding issues.

Another interesting frustrated two-dimensional quantum spin model under our consideration is the checkerboard antiferromagnet.
Similar to the case of $J_1$-$J_2$ model, we can simulate the checkerboard antiferromagnet
model by using ultracold fermionic atoms confined in the checkerboard optical
lattice. The model Hamiltonian contains both the nn couplings $J_1$ and the nnn diagonal link $J_2$.
In the limit of $J_2/J_1 \ll 1$,
its ground state has the N\'eel long-range order,  whereas it corresponds decoupled
Heisenberg chains when $J_2/J_1 \gg 1$.
According to previous studies~\cite{Starykh, Sindzingre},
there may exist three distinct quantum phases: the N\'eel order shows up for $J_2/J_1< 0.8$,
a valence bond crystal in singlet plaquettes may appear when $0.8<J_2/J_1<1.1$, and
decoupled Heisenberg spin chains corresponds to the parameter region $ J_2/J_1 > 1.1$.
The checkerboard antiferromagnet model in the strongly frustrated regime  $ J_2/J_1 \sim 1$
can be realized experimentally with ultracold atoms in optical lattices.

\section{Detecting the quantum many-body states}

Besides the preparation of the two-dimensional frustrated spin models in optical lattices,
another crucial procedure is to detect faithfully the quantum many-body states in these strongly correlated
systems. Among many different experimental methods,
one of the significant progresses in this field is the quantum polarization
spectroscopy of the atomic ensemble, which is a type of quantum
non-demolition detection~\cite{Eckert}. After a polarized probe
laser traveling through the atomic ensemble, the spin fluctuation and
correlation of the atomic system will be mapped on those of light, which can then be efficiently
measured using homodyne detection.
This method can be used to probe the spatially resolved spin-spin correlation functions and then
may distinguish different magnetic quantum phases.
Recently, this method has been applied to study the phase diagram of spin-1 bilinear-biquadratic model and appears to be quite successful~\cite{Chiara}.
In the following, we will employ the quantum polarization
spectroscopy method to detect the quantum many-body states of a small-scale $J_1$-$J_2$ model and gain some
important insights on distinct quantum phases and quantum phase transition points.

\begin{figure}
\begin{center}
\includegraphics[height=1.8in,width=2.4in]{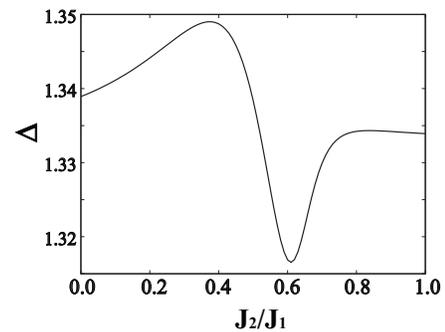}
\caption{The noise signal $\Delta$ as a function of $J_2/J_1$ of the $J_{1}$-$J_{2}$
frustrated spin model.}
\end{center}
\end{figure}

Next we consider a probe standing wave polarized in the $x$ direction
characterized by the Stokes operators $\hat s_{1}$, $\hat s_2$,
$\hat s_3$. We define a macroscopic Stokes operator $\hat S_i
=\int \hat s_i dt$ and introduce the canonical quadrature
operators $\hat X=\sqrt{2/{N_P}} \hat S_2$ , where $N_P$ is the
number of photons. The variance of $\hat X$ of the output light
contains the atomic spin-spin correlations:
\begin{equation}
\langle{(\Delta \hat X_{
out})^2}\rangle=\frac12+\frac{\kappa^2}{FN_A}\sum_{k,l=1} c_k c_l
(\langle \hat S^k_z \hat S^l_z\rangle-\langle \hat S^k_z \rangle
\langle \hat S^l_z\rangle),
\end{equation}
where $\kappa$ is the coupling constant, $F$ is the total angular
momentum of a atom, $N_A$ is the total number of atoms, $\hat S^k_z$
denotes the $z$ component spin operator of the $k$th atom. $c_n$
describes the atom-light coupling and is given by
\begin{equation}
c_n(k_p,b)=2\int dr\cos^2(k_p(r-b))|w(r-r_n)|^2,
\end{equation}
where $k_p$ is the wavevector of the probe laser, $b$ is its shift
with the optical lattice and $w(r-r_n)$ is the Wannier
wavefunction of the atom confined at site $r_n$.
Here we choose $F=1/2$ and  $\kappa=1$. The probe laser
propagates along the diagonal direction with $k_p=k/4 $ and $b=0$,
so the modification of the atom-light coupling $c_n$ has the
structure
\begin{equation}
\begin{pmatrix}
2 &  1 & 0 &  1 \\
1 & 0& 1 &  2 \\
0 &  1 & 2 & 1 \\
1&  2 &   1 & 0
\end{pmatrix}.
\end{equation}

The noise signal is defined as $\Delta=\langle{(\Delta \hat X_{
out})^2}\rangle-1/2$, which is a linear combination of various spatially varying correlation functions.
By tuning the parameter like $J_2/J_1$ of the Hamiltonian, the ground state wavefunction may evolve accordingly.
When the tuning parameter crosses the quantum phase transition point, a sudden change of the symmetry of ground state wavefunction may occur.
The quantum phase transition may lead to the possible discontinuity of correlation function itself or its first order derivative.
We naturally expect that the features of noise signal as a function of parameter should
have certain connections with distinct quantum phases and quantum phase transition points.
Then we numerically investigate the evolution of
$\Delta$ as a function of parameter $J_2/J_1$ of the frustrated
spin models and expect to reveal the connections between noise signal and quantum phase transitions.

\begin{figure}
\begin{center}
\includegraphics[height=1.5in,width=3.2in]{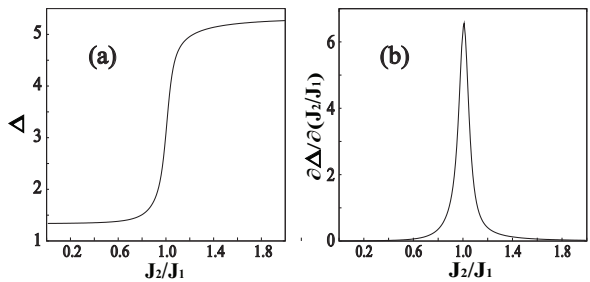}
\caption{(a) $\Delta$ as a function of $J_2/J_1$ of the checkerboard antiferromagnet model;
 (b) Its first derivative versus $J_2/J_1$. }\label{fig4}
\end{center}
\end{figure}

We numerically calculate the ground state many-body wave function
of the spin model by exact diagonalization method for a small cluster~\cite{Chen}.
Then we compute the spatially varying spin-spin correlation functions
and finally we obtain the value of the noise signal.
For the $J_1$-$J_2$ frustrated model, the noise signal $\Delta$ is
plotted as a function of $J_2/J_1$ in Fig. 3. It can be clearly seen
that there are two local extreme points at $J_2/J_1 \sim 0.38$ and $0.62$, respectively.
By analyzing the first derivative of noise signal $\Delta$ as a function of $J_2/J_1$, we may locate another
extremum point at $J_2/J_1 = 0.5$.
According to our natural expectation, we may connect these three extreme points to three possible
quantum phase transition points.
Very interestingly, these extreme points coincide quite
precisely with the known quantum phase transition points (in an infinite
system) respectively at $(J_2/J_1)_c=0.38$, $0.5$ and $0.6$~\cite{Capriotti, Singh, Sushkov}.
This one-to-one correspondence between the extreme points of noise signal and
quantum phase transition points is remarkable.
According to previous studies, two points at $(J_2/J_1)_c=0.38$ and $0.6$ are connected to
the second order quantum phase transition while the point at  $(J_2/J_1)_c=0.5$ may correspond to first order quantum phase transition.
That may explain why
the two extreme points show up for $\Delta$ itself while one extremum point appears for the first derivative of  $\Delta$.
In view of the small 4$\times$4 cluster used in our calculation, the good agreement of our results
with previous studies is truly impressive.
In other words, we demonstrate that the polarization spectroscopy method can be used to determine
the quantum phase transition points straightforwardly and characterize distinct quantum phases.

For the checkerboard antiferromagnet model, we calculate the noise signal $\Delta$ and its first derivative with respect to
$J_2/J_1$. As shown in Fig. 4, there appears no extreme point for $\Delta$ itself, the evolution of $\Delta$ as a
function of  $J_2/J_1$ can be roughly classified into three different regions, two flatter regions for both $J_2/J_1 < 0.8$ and $J_2/J_1 > 1.2$,
and one steeper region for $0.8 < J_2/J_1 < 1.2$.
By analyzing the first derivative  $\Delta$ as a function of $J_2/J_1$, there is a clear maximum point around $J_2/J_1=1$.
From the previous studies~\cite{Starykh, Sindzingre}, there might be three different phases dividing by two quantum phase transitions points
which are located at $(J_2/J_1)_c \sim 0.8$ and $1.1$, respectively. In this case, there is no good one-to-one correspondence between the
extreme points of noise signal and the quantum phase transition points. Since the unit cell of checkerboard lattice
is larger than that of simple square
lattice, a larger cluster is required to do numerical calculations.
Unfortunately, it is now beyond our computational capabilities.

\section{Summary}

Simulation of geometrically frustrated quantum antiferromagnets of ultracold fermionic atoms in optical lattices
and detection of many-body states become a very hot topic both experimentally and theoretically.
In this paper, we have shown such strongly correlated systems can be simulated by a specific optical engineering technique. In particular,
both the frustrated $J_1$-$J_2$ Heisenberg model and checkerboard antiferromagnet model in the strongly frustrated regime can be realized for
ultracold fermionic atoms in optical lattices.
Moreover we employ the polarization spectroscopy method to probe the quantum many-body states of such systems. This method is
 designed to detect the spatially varying spin-spin correlation functions. In particular, for the  $J_1$-$J_2$ model,
the quantum phase transition points can be determined straightforwardly and distinct quantum phases can be well characterized.
The polarization spectroscopy method is quite promising to probe various magnetic orders of geometrically frustrated systems.

\section{Acknowledgments}
This work was funded by the National Basic Research Program of China (Grants
No. 2009CB929204, No. 2012CB921604, and No.2011CB921204), the National Natural Science Foundation of China (Grants
No. 11174270, No. 11004186, No. 10874032, No. 10875110, No. 11074043, No. 60836001, and No. 60921091), the Research
Fund for the Doctoral Program of Higher Education of China (Grant No. 20103402110024), the Fundamental Research Funds
for the Central Universities (Grant No. WK2470000006), and the Shanghai Municipal Government. H.M. is grateful for the
support from the Fudan Undergraduate Research Opportunities Program and the National Science Fund for Talent Training
in Basic Science. Z.-W.Z. gratefully acknowledges the support of the K. C. Wong Education Foundation, Hong Kong¡£

\end{document}